\documentclass[osajnl,twocolumn,showpacs,superscriptaddress,10pt]{revtex4-1} 
\usepackage{amsmath,amssymb,graphicx}
\begin{document}

\title{Multiscale Talbot effects in Fibonacci geometry}

\author{I-Lin Ho}
\email{Corresponding author: sunta.ho@msa.hinet.net}
\affiliation{Department of Physics, National Cheng Kung University, Tainan 701, Taiwan, R.O.C.}
\author{Yia-Chung Chang}
\affiliation{Research Center for Applied Sciences, Academia Sinica, Taipei 115, Taiwan, R.O.C.}

\begin{abstract}
This article investigates the Talbot effects in Fibonacci geometry by introducing the cut-and-project construction, which allows for capturing the entire infinite Fibonacci structure into a single computational cell. Theoretical and numerical calculations demonstrate the Talbot foci of Fibonacci geometry at distances that are multiples $(\tau+2)(F_{\mu}+\tau F_{\mu+1} )^{-1}p/(2q)$ or $(\tau+2)(L_{\mu}+\tau L_{\mu+1} )^{-1}p/(2q)$ of the Talbot distance. Here, ($p$, $q$) are coprime integers, $\mu$ is an integer, $\tau$ is the golden mean, and $F_{\mu}$ and $L_{\mu}$ are Fibonacci and Lucas numbers, respectively. The image of a single Talbot focus exhibits a multiscale pattern due to the self-similarity of the scaling Fourier spectrum.
\end{abstract}

\ocis{(050.1950) Diffraction gratings; (070.6760) Talbot and self-imaging effects; (260.1960) Diffraction theory}

\maketitle 

The Talbot effect, also referred as self-imaging or lensless imaging, is of the
phenomena manifested by a running repetition of lateral periodic fields $u(x)$ along the wave propagation $z$ in
Fresnel diffraction. The simplicity and beauty of the effect triggered diverse researches related to coherent optical signal processing, and resulted in numerous interesting and original applications\cite{review1,review2,review3}, providing
competitive solutions to scientific and technological problems\cite{pb1,pb2}.

On the other hand, Montgomery\cite{proven1} demonstrated that lateral periodicity is a sufficient, but not a necessary, condition for self-imaging, and hence encouraged unexplored avenues to relevant fundamental optical sciences and device technology. In particular, aperiodic optical media generated by mathematical
rules recently attracted significant attention in
optics communities due to their unconventional nature\cite{qc1,review4} and full compatibility with current
materials deposition and device technologies\cite{tec1,tec2,tec3}. Several recent researches have leveraged
on aperiodicity as a novel strategy to engineer optical manipulation, devices, and functionalities\cite{ap1,ap2}.

\begin{figure}[htbp]
\centerline{\includegraphics[width=0.85\columnwidth]{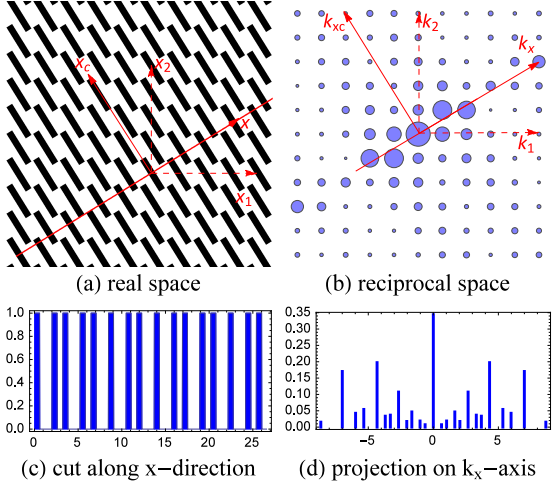}}
\caption{ (a) indicates the cut operation to obtain the transmittance function of a Fibonacci structure from a
two-dimensional periodic array of strips with $w=\Lambda/4=1/2$ $\mu m$; (b) indicates the equivalent projection
operation in its reciprocal space; a bubble chart with
bubble size proportional to the amplitude of Fourier coefficients; (c) shows the generated transmittance function $u(x)$ from (a); and (d) shows the Fourier spectrum amplitudes $|\bar{u}_{mn,k_{x}}|$ from (b) for Fibonacci structure.}\label{fig1}
\end{figure}

This present work proposes to transfer the concept of self-imaging to lateral-aperiodic objects\cite{qc2,qc3}, and mainly investigates the Talbot effects for Fibonacci geometry.
Considering the Talbot effects are
distinct only in the paraxial approximation and when the illuminated geometry tends to infinity, this work introduces
the cut-and-project construction techniques\cite{qc1,qc5}, which capture the entire infinite Fibonacci structure in
a single two-dimensional computational cell. The cut-and-projection procedure illustrates the fact that the Fourier transform of a projection operation is equivalent to a cut operation and vice versa. In Figure \ref{fig1}(a), the transmittance function of Fibonacci structure $u(x)$ along the dimension $x$ is obtained as an irrational cut of a periodic array of strips in the two-dimensional space $(x_1,x_2)$. This leads to a result of the Fourier spectrum $\bar{u}_{mn,k_{x}}$ becoming a function of the projection of the two-dimensional reciprocal lattice $(m\vec{k}_{x1}+n\vec{k}_{x2})$ on the $k_x$ axis as in Figure \ref{fig1}(b). Herein, the strips with width $w\left( x\right)$ and length $\Delta\left( x_c\right)=\Lambda\left[ \cos(\alpha)+\sin(\alpha)\right]$ are rotated by an angle $\alpha=\tan^{-1}\left[ 1/\tau\right]$ with respect to the basis $\left( x_1,x_2\right)$ of the lattice having periodicity $\Lambda$. $\vec{k}_{x1}$ and $\vec{k}_{x2}$ are the basis vectors in the reciprocal space, and $\tau=(1+\sqrt{5})/2$ is the golden mean.

\begin{figure}[htbp]
\centerline{\includegraphics[width=0.95\columnwidth]{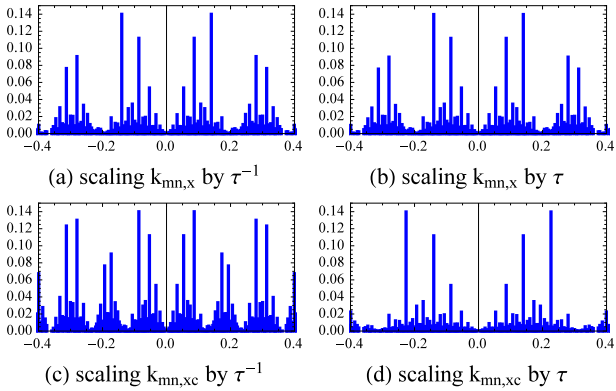}}
\caption{Fourier spectrum amplitudes $|\bar{u}_{mn,k_{x}}|$ along the $k_x$ axis with (a) scaling $k_{mn,x}$ by $\tau^{-1}$, (b) scaling $k_{mn,x}$ by $\tau$, (c) scaling $k_{mn,xc}$ by $\tau^{-1}$, and (d) scaling $k_{mn,xc}$ by $\tau$. Relevant parameters are set by $w=\Lambda/100=1$ $\mu m \ll \Lambda$.  }\label{fig2}
\end{figure}

The transmittance function $u\left( x,z=0\right)$ for a Fibonacci structure at $%
z=0$ can be given by\cite{qc1}
\begin{eqnarray}
u\left( x,z=0\right)  &=&\sum_{m,n=-\infty }^{\infty }\bar{u}_{mn,k_{x}}\exp
\left( ik_{mn,x}\cdot x\right)   \label{ux} \\
\bar{u}_{mn,k_{x}} &=&\frac{2}{\pi }\frac{\sin \left( \frac{k_{mn,x}w}{2}%
\right) \sin \left( \frac{k_{mn,xc}\Delta }{2}\right) }{k_{mn,x}w\cdot
k_{mn,xc}\Delta }  \label{uk} \\
k_{mn,x} &=&\frac{2\pi }{\Lambda }\frac{\left( m+n\tau \right) }{\sqrt{%
2+\tau }};k_{mn,xc}=\frac{2\pi }{\Lambda }\frac{\left( n-m\tau \right) }{%
\sqrt{2+\tau }}  \label{kxc}
\end{eqnarray}
It is easy to check that the distribution of $(L,S)$ segments of the function $u\left( x,z=0 \right)$ in Eq. (\ref{ux}) obeys a Fibonacci sequence $LSLSLLSLLSLSLLS...$, in which $L=\Lambda\cos\left(\alpha\right)$ and $S=\Lambda\sin\left(\alpha\right)$. This can also be seen in Figure \ref{fig1}(c). Other properties of the Fibonacci Fourier transform are that, with the relation $\tau^{\mu+1}=\tau^{\mu}+\tau^{\mu-1}$, the sequence of Fourier vectors defined by Eq. (\ref{kxc}) can be shown to be invariant when multiplied by any power of $\tau$ \cite{qc1}. Moreover, in the limit of infinitesimal strip width $w\ll \Lambda$, the Fourier spectrum $\bar{u}_{mn,k_{x}}$ in Eq. (\ref{uk}) keeps invariant after scaling $k_{mn,x}$ by $\tau^{\mu}$, since $k_{mn,x}\rightarrow \tau^{\mu} k_{mn,x}$ gives $\bar{u}_{mn,k_{x}}\left( k_{mn,x}\right)\rightarrow \bar{u}_{mn,k_{x}}\left( \tau^{\mu} k_{mn,x}\right)=\bar{u}_{mn,k_{x}}\left( k_{mn,x}\right)$. The Fourier spectrum however is rescaled linearly along $k_{x}$ axis after scaling $k_{mn,xc}$ by $\tau^{\mu}$, since $k_{mn,xc}\rightarrow \tau^{\mu} k_{mn,xc}$ leads to the approximation $\bar{u}_{mn,k_{x}}\rightarrow \bar{u}_{[\tau ^{\mu }m][\tau ^{\mu }n],k_{x}}$ mathematically. Note that $\tau$, $m$, and $n$ should keep finite in the paraxial approximation. Figure \ref{fig2} numerically shows the self-similarity of the Fourier spectrum amplitude with different $k_{mn,x}$ or $k_{mn,xc}$ scalings.

\begin{figure}[htbp]
\centerline{\includegraphics[width=0.96\columnwidth]{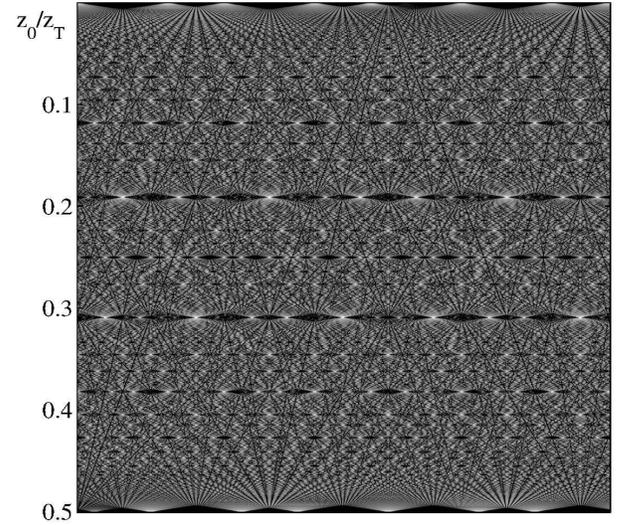}}
\caption{Talbot self-imaging for Fibonacci geometry with parameters $w=\Lambda/100=1$ $\mu m$, $\lambda=0.1\mu m$, and $|m,n|\leq 30$. The figure shows the distribution of $|u_{2d}(x,z_{0})|^{2}$ within the propagation range $0\leq z_{0}\leq z_{T}/2$ only for clarity purpose. The profile within the other half-period $z_{T}/2\leq z_{0}\leq z_{T}$ can be obtained by present figure with the relation $|u_{2d}(x,z_{T}-z_{0})|^{2}=|u_{2d}(x,z_{0})|^{2}$ }\label{fig3}
\end{figure}

We can calculate the propagation field $u\left( x,z_{0}\right) $ with wavelength $\lambda $ at
distance $z=z_{0}$ by the method of the angular spectrum
of plane wave\cite{frac1}. In Fresnel diffraction, it is given that%
\begin{eqnarray}
u_{2d}\left( x,z_{0}\right)  &=&\sum_{m,n=-\infty }^{\infty }\bar{u}%
_{mn,k_{x}}\exp \left( ik_{mn,x}\cdot x\right)   \nonumber \\
&&\cdot \exp \left[ -i\left( k_{mn,x}^{2}+k_{mn,xc}^{2}\right) \lambda
z_{0}/\left( 4\pi \right) \right]   \label{u2d} \\
u_{1d}\left( x,z_{0}\right)  &=&\sum_{m,n=-\infty }^{\infty }\bar{u}%
_{mn,k_{x}}\exp \left( ik_{mn,x}\cdot x\right)   \nonumber \\
&&\cdot \exp \left[ -i\left( k_{mn,x}^{2}\right) \lambda z_{0}/\left( 4\pi
\right) \right]   \label{u1d}
\end{eqnarray}
Here, the field $u_{2d}\left( x,z_{0}\right)$ represents the measurement of a $x$-cut-line of the two-dimensional illumination in Figure (\ref{fig1}a), and the field $u_{1d}\left( x,z_{0}\right)$ is for the measurement of a one-dimensional grating in Figure (\ref{fig1}c).
It is straightforward to define the primary Talbot distance $z_{T}=2\Lambda^{2}/\lambda$ by the relation $k^{2}_{mn,x}+k^{2}_{mn,xc}=4\pi^{2}\left(m^{2}+n^{2}\right)/\Lambda^{2}$ for the field $u_{2d}$. Since the dominant spots of $\bar{u}_{mn,k_{x}}$ occur with $k_{mn,xc}$ having small $n-m\tau$ (see Eqs. (\ref{uk})-(\ref{kxc})), the function $u_{1d}$ can be treated as a perturbation case of the function $u_{2d}$ with a phase deviation $\exp\left[ ik^{2}_{mn,xc}\lambda z_{0}/(4\pi)\right]$ at small $z_{0}<z_{T}$. The properties of $u_{1d}$ will be discussed in the last paragraph. Applying Eq. (\ref{u2d}), the Talbot self-imaging for Fibonacci geometry is calculated and shown in Figure \ref{fig3}.

\begin{figure}[htbp]
\centerline{\includegraphics[width=0.96\columnwidth]{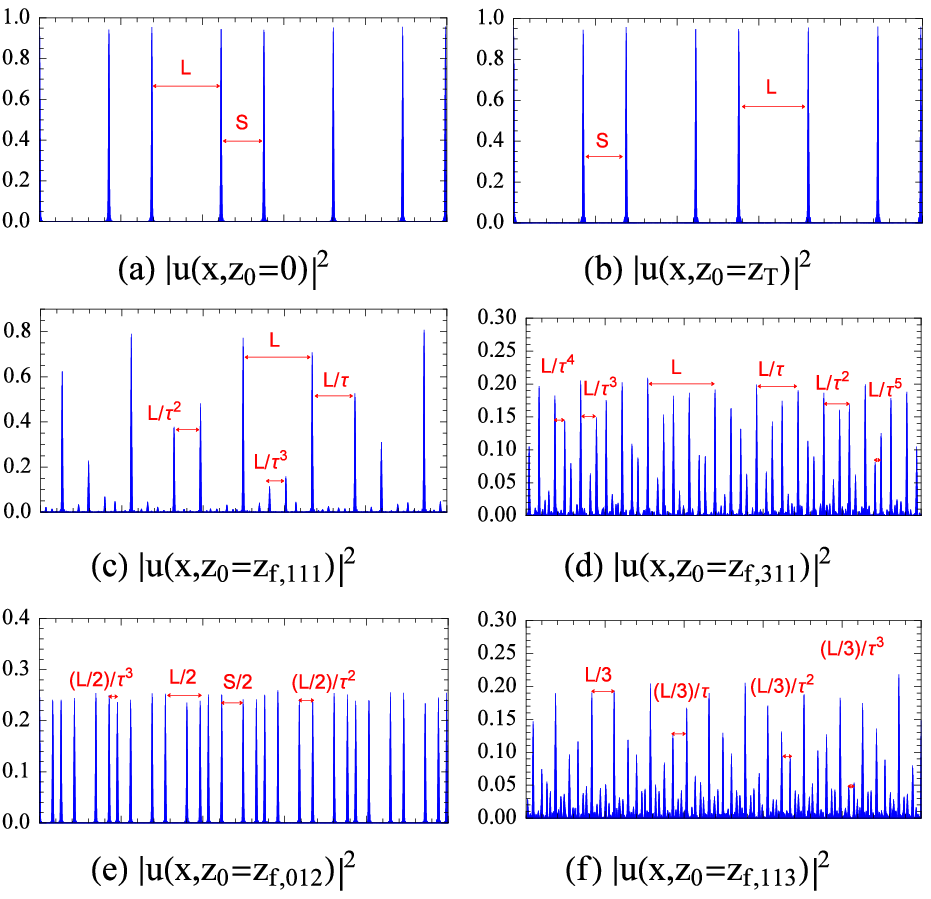}}
\caption{ Talbot images at (a) $z_{0}=0$, (b) $z_{0}=z_{T}$, (c) $z_{0}=z_{f,111}$, (d) $z_{0}=z_{f,311}$, (e) $z_{0}=z_{f,012}$, and (f) $z_{0}=z_{f,113}$. Parameters are defined the same as that in Figure \ref{fig3}.} \label{fig4}
\end{figure}

It can be seen that, aside from the conventional fractional Talbot foci at $z_{0}=z_{T}\cdot p/q$\cite{frac1}, Figure \ref{fig3} shows the irrational Talbot effects as well. Here ($p$, $q$) are coprime integers. This also can be understood by the self-similarity of the scaling Fourier spectrum in the Fibonacci structure.
Consider the scaling of $k_{mn,x}$ by $\tau ^{g}$, the scaling of $%
k_{mn,xc}$ by $\tau ^{h}$ ($g,h$ are integers), or the linear combination of different scaling terms, e.g. $k^{2}_{mn,x}$ by $\tau^{2g+1}=\tau^{2g+2}-\tau^{2g}$, in Eq. (\ref{u2d}). Without loss of generality, the
propagation phase term $\exp \left[ -i\left(
k_{mn,x}^{2}+k_{mn,xc}^{2}\right) \lambda z_{0}/\left( 4\pi \right) \right]
\equiv \exp \left[ -i\Phi \right] $ can be given by
\begin{eqnarray}
\Phi  &=&\frac{\pi \lambda z_{0}}{\Lambda ^{2}}\left[ \frac{\left( m+n\tau
\right) ^{2}}{\tau +2}\tau ^{g}+\frac{\left( n-m\tau \right) ^{2}}{\tau +2}%
\tau ^{h}\right]   \nonumber \\
&=&\frac{\pi \lambda z_{0}}{\Lambda ^{2}}\left[ \left( m^{2}+n^{2}\right) L_{%
\frac{g-h}{2}}-\left( m^{2}-4mn-n^{2}\right) F_{\frac{g-h}{2}}\right]
\nonumber \\
&&\left[ L_{\frac{g+h}{2}}+\tau L_{\frac{2+g+h}{2}}\right] \left[ 4+2\tau %
\right] ^{-1}\text{ {\small for even} }\left( g-h\right) /2  \label{evenf} \\
&=&\frac{\pi \lambda z_{0}}{\Lambda ^{2}}\left[ 5\left( m^{2}+n^{2}\right)
F_{\frac{g-h}{2}}+\left( n^{2}+4mn-m^{2}\right) L_{\frac{g-h}{2}}\right]
\nonumber \\
&&\left[ F_{\frac{g+h}{2}}+\tau F_{\frac{2+g+h}{2}}\right] \left[ 4+2\tau %
\right] ^{-1}\text{ {\small for odd} }\left( g-h\right) /2  \label{oddf}
\end{eqnarray}%
where the relations below have been used\cite{fibonacc1,lucas1}
\begin{eqnarray}
&&F_{g+1}=\left( F_{g}+L_{g}\right) /2,\text{ }F_{g-1}=\left(
L_{g}-F_{g}\right) /2  \label{re1} \\
&&F_{g+h}+\left( -1\right) ^{h+1}F_{g-h}=F_{h}L_{g} \\
&&L_{g+h}+\left( -1\right) ^{h}L_{g-h}=L_{h}L_{g} \\
&&F_{g-h}=\left( -1\right) ^{h}\left( F_{g}L_{h}-L_{g}F_{h}\right) /2 \\
&&L_{g-h}=\left( -1\right) ^{h}\left( L_{g}L_{h}-5F_{g}F_{h}\right) /2
\end{eqnarray}%
Here, $F_{\mu }$\cite{fibonacc1} and $L_{\mu }$\cite{lucas1} are Fibonacci
and Lucas numbers, respectively. From the constraint $\Phi =\pi \ell $ ($%
\ell $ is an integer), the distance of the foci for the irrational Talbot
image can be decided by Eqs. (\ref{evenf})-(\ref{oddf})
\begin{eqnarray}
z_{f} &=&\left( 2+\tau \right) \left[ L_{\mu }+\tau L_{\mu +1}\right] ^{-1}%
\frac{z_{T}}{2}\text{ \ {\small for even} }\frac{g-h}{2}  \label{even2} \\
&=&\left( 2+\tau \right) \left[ F_{\mu }+\tau F_{\mu +1}\right] ^{-1}\frac{%
z_{T}}{2}\text{ \ {\small for odd} }\frac{g-h}{2}  \label{odd2}
\end{eqnarray}%
Here, $\mu =(g+h)/2$ is an integer. It is easy to show that $z_{f}$ is equal
to the conventional Talbot distance $z_{T}/2$ \cite{qcf} for $g=h=0$. Note
that the coefficients in the first square bracket in Eqs. (\ref{evenf}) and (%
\ref{oddf}) give an even integer, which can be proven by Eq. (\ref{re1}).
Consequently an extra factor $1/2$ entered Eqs. (\ref{even2}) and (\ref{odd2}%
). Together with the extension of the fractional Talbot effect\cite{frac1}
for Fibonacci geometry, the foci of the function $u_{2d}$ can be decided at
the distance
\begin{eqnarray}
z_{f,\mu pq} &=&\frac{z_{T}}{2}\left( 2+\tau \right) \left[ L_{\mu }+\tau
L_{\mu +1}\right] ^{-1}pq^{-1}\text{\ }  \nonumber \\
&&\text{ \ \ \ \ \ \ \ \ \ \ \ \ \ \ \ \ \ \ \ \ \ \ \ \ \ for even }(g-h)/2
\label{even4} \\
&=&\frac{z_{T}}{2}\left( 2+\tau \right) \left[ F_{\mu }+\tau F_{\mu +1}%
\right] ^{-1}pq^{-1}  \nonumber \\
&&\text{ \ \ \ \ \ \ \ \ \ \ \ \ \ \ \ \ \ \ \ \ \ \ \ \ \ for odd }(g-h)/2
\label{odd4}
\end{eqnarray}
with additional conditions of $|u_{2d}(x,z_{f,\mu pq}+\ell
z_{T})|^{2}=|u_{2d}(x,z_{f,\mu pq})|^{2}$ and $|u_{2d}(x,z_{T}-z_{f,\mu
pq})|^{2}=|u_{2d}(x,z_{f,\mu pq})|^{2}$. Remind that the scaling of $k_{mn,x}$ ($\propto\tau^{g}$) makes the sequence of Fourier vectors and the Fourier spectrum $\bar{u}_{mn,k_{x}}$ invariant, and so dose the term $\bar{u}_{mn,k_{x}}\exp\left( ik_{mn,x}\cdot x\right)$ in Eq. (\ref{u2d}). The scaling of $k_{mn,xc}$ ($\propto\tau^{h}$), however, signifies a rescale of $\bar{u}_{mn,k_{x}}$ along the $k_{x}$ axis (see Figures \ref{fig2}), and equivalently suggests an inverse scaling of $u_{2d}(x,z_{0})$ along $x$ axis by $\bar{u}_{[\tau ^{\mu }m][\tau ^{\mu }n],k_{x}}\exp \left( ik_{mn,x}\cdot
x^{\prime }\right) \rightarrow \bar{u}_{mn,k_{x}}\exp \left( ik_{mn,x}\cdot
x^{\prime }\tau ^{-\mu }\right) \equiv \bar{u}_{mn,k_{x}}\exp \left(
ik_{mn,x}\cdot x\right) $ in Eq. (\ref{u2d}). Since $\mu=(g+h)/2$,
the function $z_{f,\mu pq}$ hence determines the Talbot foci of Fibonacci geometry having multiscale segments of $(L/q)\tau^{-\nu}$ or $(S/q)\tau^{-\nu}=(L/q)\tau^{-\nu-1}$ with $\nu\in\{0,1,...,|\mu|\}$ in rough. Figure \ref{fig4} depicts the numerical results for several $z_{f,\mu pq}$, and presents multiscale Talbot effects for Fibonacci geometry.
Figures \ref{fig4}(a) and \ref{fig4}(b) show the origin and the repetition of the Fibonacci structure at $z_{0}=0$ and $z_{0}=z_{T}$, respectively. Figure \ref{fig4}(c) is the focus image at $z_{0}=z_{f,111}=0.3090z_{T}$ having a dominant $\tau^{-1}$ and $\tau^{-2}$ scaling pattern. Figure \ref{fig4}(d) is the focus image at $z_{0}=z_{f,311}=0.2639z_{T}$ showing the distinct scaling pattern up to the order about $\tau^{-5}$. Figure \ref{fig4}(e) is the fractional image at $z_{0}=z_{f,012}=0.25z_{T}$ illustrating half-length Fibonacci segments $L/q=L/2$ and $S/2$. Figure \ref{fig4}(f) is the focus image at $z_{0}=z_{f,113}=0.1030z_{T}$ depicting the $1/3$-shrinking as well as the $\tau$-scaling configurations.

\begin{figure}[htbp]
\centerline{\includegraphics[width=0.96\columnwidth]{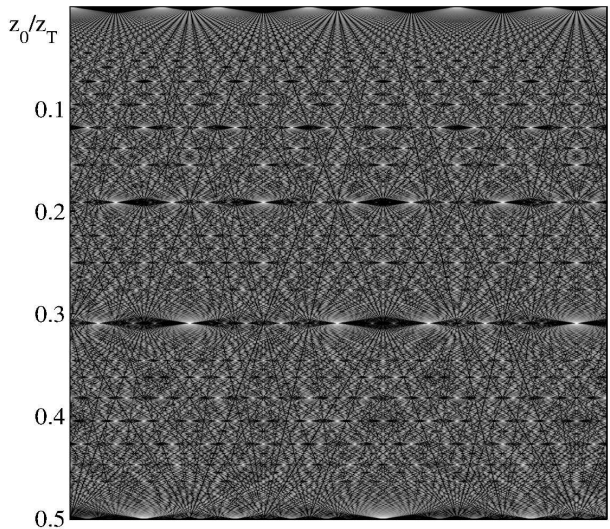}}
\caption{Talbot self-imaging for Fibonacci geometry with parameters $w=\Lambda/100=1$ $\mu m$, $\lambda=0.1\mu m$, and $|m,n|\leq 30$. The figure shows the distribution of $|u_{1d}(x,z_{0})|^{2}$ within the propagation range $0\leq z_{0}\leq z_{T}/2$. }\label{fig5}
\end{figure}

Figure \ref{fig5} shows the Talbot self-imaging of Fibonacci structure $u_{1d}$ using the same parameters for Figure \ref{fig3}. This image of $u_{1d}$ presents similar properties to that of $u_{2d}$ (see Figure \ref{fig3}) at $z_{0}<z_{T}$ except for small perturbations of the term $\exp\left[ ik^{2}_{mn,xc}\lambda z_{0}/(4\pi)\right]$ by comparing Eqs. (\ref{u2d}) and (\ref{u1d}). With the increasing propagation distance $z_{0}>z_{T}$, the phase deviation is amplified and it raises a significant departure from the inference by Eqs. (\ref{even4})-(\ref{odd4}). In fact, the ergodicity of phase of $u_{1d}$ at $z_{0}\gg z_{T}$ will be unsettled until further studies.

To conclude, this article investigates the Talbot properties for Fibonacci geometry constructed by the cut-and-projection method. Analytical formulae for Talbot foci are deduced, which suggest for two feasible experiments. Theoretical and numerical calculations demonstrate fractional and irrational Talbot images, and exhibit a multiscale Talbot effect due to the self-similarity of the scaling Fourier spectrum in the Fibonacci structure.

This work was supported in part by National Cheng-Kung University, National Center for Theoretical Science (South), and the National Science Council of the Republic of China under Contract Nos.
NSC 101-2112-M-001-024-MY3.

\end{document}